\documentclass[12pt,a4paper]{article}

\usepackage[british]{babel}

\usepackage[a4paper,top=2cm,bottom=2cm,left=2.5cm,right=2.5cm,marginparwidth=1.75cm]{geometry}

\usepackage[style=apa, backend=biber]{biblatex} 
\DeclareLanguageMapping{british}{british-apa} 
\DeclareFieldFormat[article]{volume}{\apanum{#1}} 

\usepackage{amsmath}
\usepackage{graphicx}
\usepackage[colorlinks=true, allcolors=blue]{hyperref}
\usepackage{hyperref}
\usepackage{orcidlink}
\usepackage[title]{appendix}
\usepackage{mathrsfs}
\usepackage{amsfonts}
\usepackage{booktabs} 
\usepackage{caption}  
\usepackage{threeparttable} 
\usepackage{algorithm}
\usepackage{algorithmicx}
\usepackage{algpseudocode}
\usepackage{listings}
\usepackage{enumitem}
\usepackage{chngcntr}
\usepackage{booktabs}
\usepackage{lipsum}
\usepackage{subcaption}
\usepackage{authblk}
\usepackage[T1]{fontenc}    
\usepackage{csquotes}       
\usepackage{diagbox}

\usepackage{setspace}
\usepackage{titlesec}
\titleformat{\section} 
  {\normalfont\Large\bfseries}{\thesection.}{1em}{}
  
\usepackage{lineno} 

\rightlinenumbers 

\linenumbers 

\usepackage{float}   
\usepackage{caption} 


\title{Optimization of Functional Materials Design with Optimal Initial Data in Surrogate-Based Active Learning}

\author[1*]{Seongmin Kim}
\author[1*]{In-Saeng Suh}
\affil[1]{\small Oak Ridge National Laboratory, National Center for Computational Sciences, Oak Ridge, 37830, TN, USA}

\affil[*]{Corresponding author: \texttt{kims@ornl.gov}, \texttt{suhi@ornl.gov}}

\date{}  

\begin{document}
\maketitle

\begin{abstract}
The optimization of functional materials is important to enhance their properties, but their complex geometries pose great challenges to optimization. Data-driven algorithms efficiently navigate such complex design spaces by learning relationships between material structures and performance metrics to discover high-performance functional materials. Surrogate-based active learning, continually improving its surrogate model by iteratively including high-quality data points, has emerged as a cost-effective data-driven approach. Furthermore, it can be coupled with quantum computing to enhance optimization processes, especially when paired with a special form of surrogate model ($i.e.$, quadratic unconstrained binary optimization), formulated by factorization machine. However, current practices often overlook the variability in design space sizes when determining the initial data size for optimization. In this work, we investigate the optimal initial data sizes required for efficient convergence across various design space sizes. By employing averaged piecewise linear regression, we identify initiation points where convergence begins, highlighting the crucial role of employing adequate initial data in achieving efficient optimization. These results contribute to the efficient optimization of functional materials by ensuring faster convergence and reducing computational costs in surrogate-based active learning. 
\end{abstract}

\textbf{Keywords}: active learning, surrogate model, factorization machine, optimization, machine learning, initial data


\section{Introduction}

The optimal design of functional materials has become increasingly essential for enhancing their properties \cite{molesky2018inverse, zunger2018inverse, liu2020machine}. However, their inherently complex geometrical features significantly expand the design spaces, posing challenges to optimization processes \cite{himanen2019data, kitai2020designing, kusne2020fly}. Exploring such large design spaces is experimentally and computationally expensive, thus several optimization algorithms have been proposed to tackle these challenges, such as neural network, Bayesian optimization, genetic algorithm, and black box model \cite{chen2020generative, shang2023hybrid, wei2020genetic, jiang2021metasurface}. These data-driven algorithms aim to learn the underlying relationships and patterns between material structures and their corresponding performance metrics (i.e., figure of merit; FOM) within available data, enabling the design of high-performing functional materials through making informed decisions. These algorithms have found successful applications across a wide range of domains, such as batteries, thermoelectric materials, metamaterial optical materials, and photonic materials \cite{liu2020machine, wang2020machine, ha2023rapid, ma2021deep}. These applications strongly demonstrate the advantages offered by data-driven approaches in optimizing functional materials to achieve desired properties.

Surrogate-based active learning approaches iteratively build a surrogate model and add a higher quality data point (i.e., a pair of the material structure and the corresponding performance) to the previous dataset after a decision-making process \cite{lye2021iterative, pestourie2020active, kapadia2024active}. These techniques have attracted a lot of interest in the field of data-driven material design over the last decade since they usually require much lower computational costs compared to other data-driven approaches \cite{pestourie2023physics, ren2021survey}. Additionally, these algorithms can be flexibly integrated with quantum computing, offering significant acceleration for the decision-making process \cite{wilson2021machine, kitai2020designing, kim2024performance, kim2024wide}. Here, quantum computing has shown great promise in exploring optimization spaces when a surrogate model is translated into a quadratic unconstrained binary optimization (QUBO) formulation \cite{pastorello2019quantum}. Kiati et. al. and Kim et al. demonstrated that quantum computing-enhanced active learning schemes significantly speed up optimization processes on quantum annealer, enabling the design of complex functional materials including radiative coolers, spectral filters, and metamaterial optical diodes \cite{kitai2020designing, kim2022high, kim2024distributed, kim2024quantum, kim2024wide}. In the schemes, factorization machine (FM), a supervised learning model describing the relationship between input vectors and corresponding output values, plays an important role. Notably, the model parameters obtained after training FM well fit the QUBO model, which means that FM can be directly connected to quantum computing without losing any information while translating surrogates from the FM model parameters into QUBOs \cite{kitai2020designing, kim2022high}. In this regard, most researchers using quantum computing-assisted active learning employ FM as the preferred machine learning model to build surrogates.

Active learning algorithms aim to converge toward an optimal state ultimately. Most studies employing active learning with FM to utilize quantum computing typically start with a fixed number of initial data points, such as 25 or 50 data, regardless of design space sizes \cite{kitai2020designing, kim2022high, kim2023design, kim2024quantum, kim2024wide}. While this approach works effectively in scenarios where the design space is relatively small, it presents challenges when the design space is large. In such cases, FM struggles to build a reliable surrogate model with limited initial data. Consequently, data pairs obtained through an optimization cycle with active learning are not necessarily of high quality; instead, they resemble randomly selected data points, generally featuring low quality. In such cases, starting optimization with an initial dataset containing more data points would be beneficial to make the algorithm capture the complexity of the optimization space early, thereby minimizing computational costs for machine learning and quantum computing. Then, the efficiency of the overall optimization process can be enhanced. 

In this work, we systematically investigate the optimal amount of initial data required to achieve faster convergence across various sizes of design spaces for surrogate-based active learning working with FM. We determine the number of iterations required for convergence across different volumes of initial datasets for varying design space sizes. Our analysis involves employing an averaged piecewise linear regression technique, which effectively captures the convergence patterns observed in scatter plots depicting FOMs as a function of optimization cycles. This regression technique proves particularly adept at accurately modeling non-smooth regression lines, effectively describing the complex distributions of FOMs. In comparison, polynomial regressions or piecewise regressions struggle to capture such complex distributions. Through this study, we offer valuable insights into determining the appropriate size of initial data required to reduce computational costs while achieving faster and more reliable convergence in the optimization process.

\section{Background}
\subsection{Surrogate-Based Active Learning}

\begin{figure}[!ht]
\centering
\includegraphics[width=1.0\linewidth]{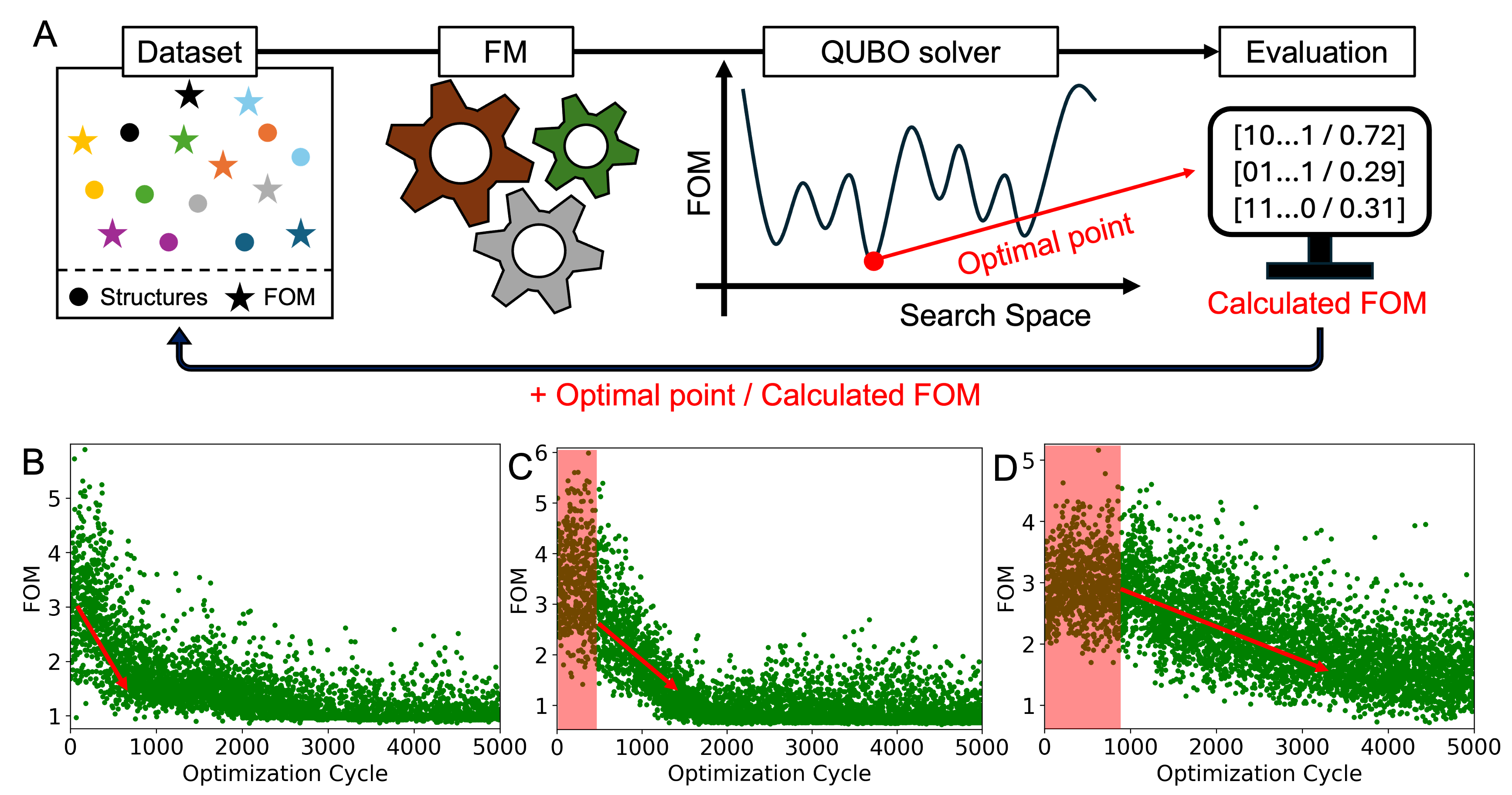}
\caption{\label{fig:fig-1} Surrogate-based active learning to optimize functional materials. (A) Schematic of the surrogate-based active learning algorithm. Optimization results after 5,000 iterations with the surrogate-based active learning algorithm for a (B) 40, (C) 60, and (D) 140-bit system.}
\end{figure}

Figure~\ref{fig:fig-1}A illustrates a workflow of the surrogate-based active learning algorithm designed to optimize functional materials through iterative processes. The active learning algorithm comprises three key components \cite{kitai2020designing, kim2022high, kim2023design, kim2024quantum, kim2024wide}: 

\textbf {(1) Factorization machine:} FM is trained with datasets to construct a surrogate model. 

\textbf {(2) Surrogate-based optimization:} The surrogate model (QUBO models), formulated with the FM model parameters, is evaluated by QUBO solvers such as quantum computing or quantum annealing to identify an approximated optimal state of the given surrogate.

\textbf {(3) Property calculation:} Functional properties associated with the approximated optimal state predicted by the QUBO solver (from step 2) are calculated.

The algorithm iteratively updates its dataset quality, which allows for building a more reliable surrogate model. Subsequently, a higher-quality data point can be identified, leading to the identification of an optimal material structure in a global design space.

\subsection{Surrogate-Based Optimization}
FM is a supervised learning algorithm that learns the relationship between input vectors \textbf{\textit {x}} and their corresponding output values $y$ with the following equation \cite{kim2022high, kim2024wide}: 
\begin{equation}
y = w_0 + \sum_{i=1}^{n} w_i {\textit{x}}_i + \frac{1}{2} \sum_{f=1}^k \left[
\left( \sum_{i=1}^n v_{i,f} {\textit{x}}_{i} \right)^2 - \sum_{i=1}^{n} v_{i,f}^2 {\textit{x}}_{i}^2 \right]
\label{eqn2}
\end{equation}
where $n$ and $k$ denote the length of the input vector \textit {\textbf {x}} and the latent space size, respectively. We set $k$ to 4 in this study. After training, FM yields model parameters ($w_0, w_i$, and $v_{i,f}$) where $w_0, w_i$, and $v_{i,f}$ represent a global bias, linear coefficient and quadratic coefficient, respectively. These model parameters can be used to build a surrogate to represent design space, describing the relationship between material structures and corresponding FOMs. 

Benefits of employing FM within the active learning algorithm include fast training, capturing complex relationships within sparse datasets, and leveraging quantum computing to handle a surrogate model derived from the FM model parameters. Here, a quantum computer takes a surrogate in the QUBO form to evaluate the energy landscape and return a binary vector that has the ground state, as the following equation \cite{kim2024quantum}:
\begin{equation}
 \bar{y} = \sum_{\textbf{\textit{x}}_{i} \in \{0,1\}^n} \textbf{\textit{x}}^T \mathbf{Q} \textbf{\textit{x}}
\label{eqn3}
\end{equation}
where \textbf {Q} and \textbf{\textit{x}} represent a QUBO matrix and binary vector, respectively. Quantum computing aims to find an optimal binary vector ($\bar{\textbf{\textit{x}}}$) that has the lowest expected output value ($\bar{y}$):
\begin{equation}
\bar{\textbf{\textit{x}}} = {\rm arg ~ \min_{\textit{x}}} ~ \bar{y}
\label{eqn4}
\end{equation}
Notably, a quantum annealer—a specialized quantum device designed for solving combinatorial optimization problems—is generally used for this task, showing great promise in identifying the ground state through the quantum adiabatic process \cite{hen2016quantum, kim2024wide}. To sum up, the FM model parameters can be directly fitted to a QUBO model, where linear and quadratic terms in a QUBO correspond to the linear and interaction coefficients of the FM model ($w$ and $v$). Consequently, a surrogate given by FM can be seamlessly linked to a QUBO, which is then solved by quantum computing (or quantum annealing) to find an optimal state \cite{kim2025distributed}. We note that classical optimization methods, such as simulated annealing (SA), can also provide good solutions, particularly when design spaces are not significantly large \cite{volpe2023integration}. 

Our goal in this study is to observe convergence patterns of FOMs, which are mainly determined by FM. Therefore, to mitigate costs associated with quantum computing, we utilize SA (D-Wave sampler) as a QUBO solver for given surrogates (i.e., QUBOs), which may yield similar optimization results to those with quantum computing, especially in these design spaces ranging from 40 to 160-bit systems.

\subsection{Optimization of Functional Materials – Transparent Radiative Cooling Window}

As a case study, we apply this active learning algorithm to design a transparent radiative cooling (TRC) window. Radiative cooling techniques, which aim at reflecting input heat sources and emitting thermal radiation through the atmospheric window (wavelength from 8 to 13 $\mu$m), have attracted considerable attention over the last decade as a solution to address the global warming issue \cite{li2019radiative, wang2021scalable, zhu2021subambient}. In particular, TRC materials can be used for building or automobile windows, which are considered the least efficient components for cooling, to minimize energy consumption \cite{dang2022ultrathin, kim2021visibly}. These TRC windows are generally designed for having high transmission in the visible regime while blocking transmission in the ultraviolet (UV) and near-infrared (NIR) regimes. Furthermore, they are designed to have high emissions in the mid to long-wave infrared regimes (M/LWIR). High emission in the M/LWIR regimes can be easily achieved by a thin polymer layer deposited on the top of TRC materials, thus a primary objective in designing TRC windows is to achieve selective sunlight transmission based on the wavelength \cite{kim2022high, kim2024wide}. 

In this work, we design planar multilayered structures for TRC windows. Each layer comprises four material candidates (silicon dioxide: SiO$_2$, silicon nitride: Si$_3$N$_4$, aluminum oxide: Al$_2$O$_3$, and titanium dioxide: TiO$_2$) with a fixed total thickness of 1,200 nm. A polydimethylsiloxane layer (40 $\mu$m thick) is deposited on the top for the emission layer, and the bottom substrate is SiO$_2$. The number of layers varies from 20 to 80. Each layer is one of the four materials, and thus it is assigned a two-digit binary label: ‘00’ for SiO$_2$, ‘01’ for Si$_3$N$_4$, ‘10’ for Al$_2$O$_3$, and ‘11’ for TiO$_2$. Therefore, 20- or 80-layered TRC windows respectively represent 40- or 160-bit systems. An ideal TRC window should exhibit unity transmission in the visible regime and zero transmission in the UV and NIR regimes. 

The objective is to design a TRC window with optical properties similar to the ideal case in terms of solar-weighted transmission, which often refers to transmitted (solar) irradiance. To evaluate the performance of TRC windows, we employ a performance metric known as the FOM, calculated using the following equation:
\begin{equation}
\text{FOM} = \frac{ 10\int_{\lambda=300}^{\lambda=2,500} [(T_{ideal}(\lambda)S(\lambda))^2 - (T_{designed}(\lambda)S(\lambda))^2]d\lambda} {\int_{\lambda=300}^{\lambda=2,500} S(\lambda)^2 d\lambda} 
\label{eqn1}
\end{equation}
where $T(\lambda)S(\lambda)$ is the transmitted irradiance, $S(\lambda)$ is the solar irradiance, $T_{designed}(\lambda)$ and $T_{ideal}(\lambda)$ indicate the transmission efficiency of a designed and ideal TRC window. Optical properties are calculated by transfer matrix method \cite{kim2022high, kim2024wide}. Note that FOM approaches 0 for higher-performance TRC windows, hence these are minimization optimization problems.

\section{Method}
\subsection{Determining Convergence}
FOM tends to decrease as the optimization cycle progresses when the active learning algorithm works well because this case (optimization of TRC windows) is designed for a minimization optimization problem (Figure ~\ref{fig:fig-1}B,C,D). To quantitatively analyze the decreasing trend of the FOM with respect to optimization cycles, we employ FOM-optimization cycle plots. First, we generate data after optimizing 40 to 160-bit TRC systems starting with different numbers of initial data (25 to 2,000). Then, we draw regression lines on the FOM-optimization cycle plots and calculate the gradients of the regression lines. As FOMs generally exhibit non-linear relationships with the optimization cycles, non-linear regression techniques such as polynomial regression or piecewise regression should be applied \cite{kim2021comparison, yang2019piecewise, jekel2019pwlf}. Several factors influence the regression plots, including the polynomial degrees, the number of pieces for piecewise regression, and the range of each piece for piecewise regression. We use polynomial degrees of 3 and 5 for polynomial regression to fit FOM distributions. Besides, we use 5 and 100 pieces with regular intervals for piecewise regression. Averaged piecewise regression takes averages of five different piecewise regressions with different ranges for the regressions, where each piecewise regression includes 20 pieces. We systematically study these regression plots with different conditions to analyze FOM convergence trends. We decide that convergence starts when the gradient of -3 in the regression line is first observed. This approach ensures a comprehensive assessment of the optimization process.

\subsection{Energy saving calculation}
Energy-saving calculations were conducted using EnergyPlus version 9.4. A standard office model with a dimension of 6 m (width) $\times$ 8 m (length) $\times$ 2.7 m (height), having two windows with 3 m (width) $\times$ 2 m (height), was considered for simulation. The model was simulated with either the optimized transparent radiative cooling (TRC) windows or conventional class windows. The target cooling temperature was set to 24$^{\circ}$C with all other default settings maintained, except for the optical properties of the optimized TRC windows (solar transmittance: 0.6650, solar reflectance: 0.3350, visible transmittance: 0.8749, visible reflectance: 0.1251, IR transmittance: 0.3860, and hemispherical emissivity: 0.5357). 
Sixteen U.S. cities (Albuquerque, Atlanta, Austin, Boulder, Chicago, Duluth, Fairbanks, Helena, Honolulu, Las Vegas, Los Angeles, Minneapolis, New York City, Phoenix, San Francisco, and Seattle) and sixteen international cities in temperate or tropical climates (Beijing, Berlin, Geneva, Incheon, London, Prague, Sapporo, Ulaanbaatar, Addis Ababa, Bangkok, Colombo, Harare, Havana, Nadi, Salvador and Singapore) were selected to calculate the energy consumption for cooling. Weather data for these cities were obtained from the EnergyPlus website.

\section{Experiments}
\subsection{FOM Convergence Analysis}
We analyze FOM convergence patterns after optimization with different initial data sizes for various design space sizes. FOM convergence can be achieved with only a few optimization cycles when starting with 25 initial data for a small design space, such as a 40-bit system (Figure~\ref{fig:fig-1}B). However, convergence requires more cycles for larger design spaces when starting optimization with the same number of initial data. For example, 60- and 140-bit systems require hundreds to a thousand optimization cycles to achieve convergence when starting with 25 initial data. Red shades in Figures~\ref{fig:fig-1}C,D indicate low-quality data (featuring high FOM) collected during early optimization cycles. These low-quality data points resemble randomly selected points, which prevent FOM from converging to optimal states. In such scenarios, it is preferable to start optimization with more initial data to achieve faster convergence, which allows us to ensure reliable optimization results and mitigate computational costs associated with FM training and surrogate solving. 

\begin{figure}[!ht]
\centering
\includegraphics[width=1.0\linewidth]{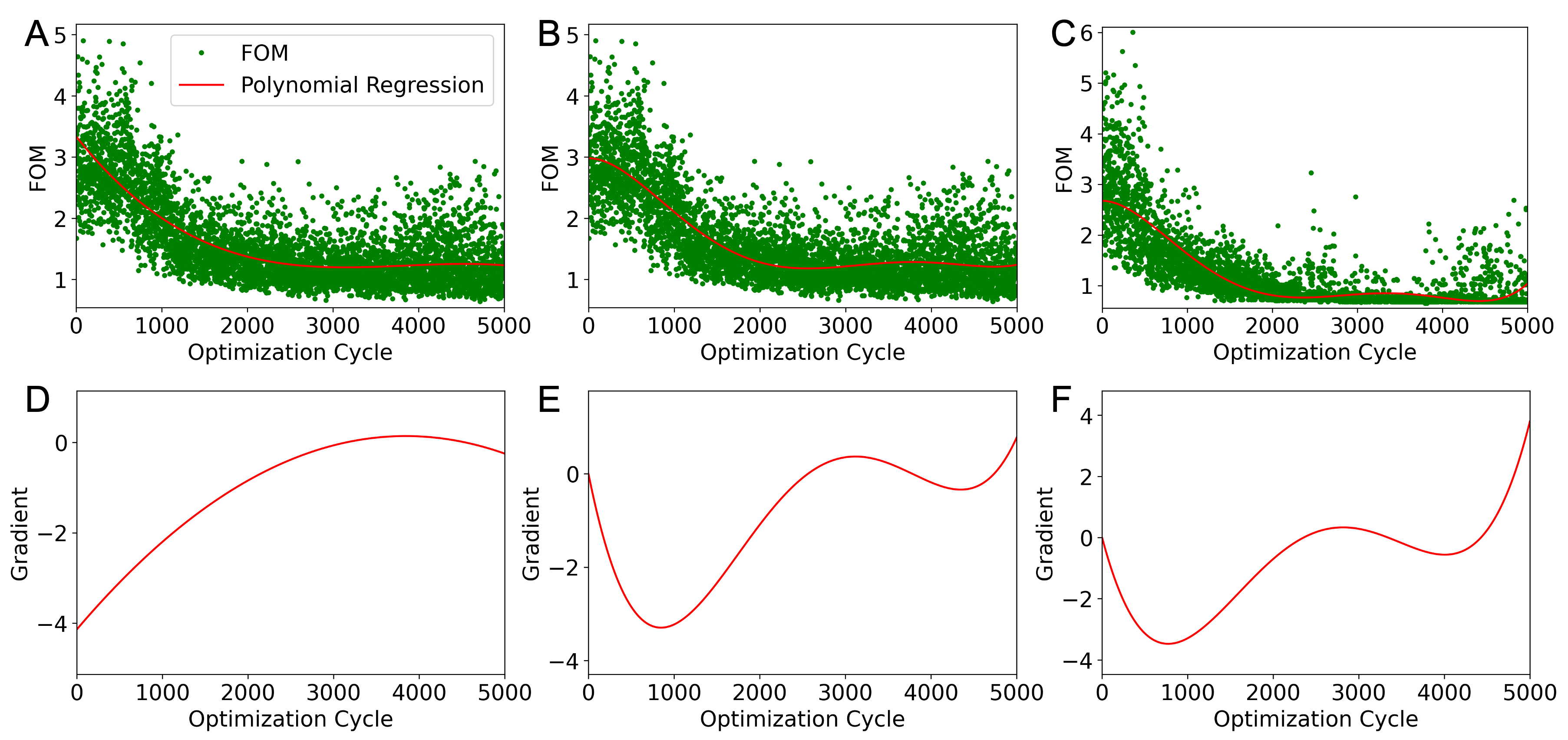}
\caption{\label{fig:fig-2} The analysis of FOM distributions (green dots) with polynomial regression (red lines). (A-B) FOM distributions and regression results after 5,000 iterations of active learning for a 120-bit system (60-layered TRC) starting optimization with 500 initial data. (C) FOM distribution and regression result after 5,000 iterations of active learning for a 40-bit system (20-layered TRC) starting optimization with 200 initial data. Polynomial regression is applied with a polynomial degree of (A) 3 and (B,C) 5. (D-F) The gradient of polynomial regression lines for Figures (A-C).}
\end{figure}

To analytically determine the initiation point where convergence starts, we calculate gradients of regression lines based on FOM-optimization cycle plots. First, we adopt a polynomial regression technique to fit the non-linear relationship between optimization cycles and corresponding FOM values. Figure~\ref{fig:fig-2}A demonstrates that polynomial regression fails to capture complex FOM distributions when a polynomial degree is low (e.g., 3). Thus, regression supposes that FOM decreases from the initial optimization cycle, resulting in a negative gradient for the regression line at the initial cycle (Figure~\ref{fig:fig-2}D). Gradient across the overall optimization features simple relation due to the low polynomial degree, which does not model the data well (Figure~\ref{fig:fig-2}D). Increasing the polynomial degree (from 3 to 5) improves the regression fit. With a higher polynomial degree, it is clear from the regression line that consistently high FOMs are observed in the early stage of optimization (until $\sim$500 cycles) and FOMs decrease after that. However, this polynomial degree cannot be universally applied to other cases. For instance, the regression with the polynomial degree of 3 does not fit well for a 120-bit system (Figure~\ref{fig:fig-2}D). The results infer that polynomial regression may not be suitable for analyzing FOM distributions due to its sensitivity to the polynomial degree \cite{gelman2019high}.

\begin{figure}[!ht]
\centering
\includegraphics[width=1.0\linewidth]{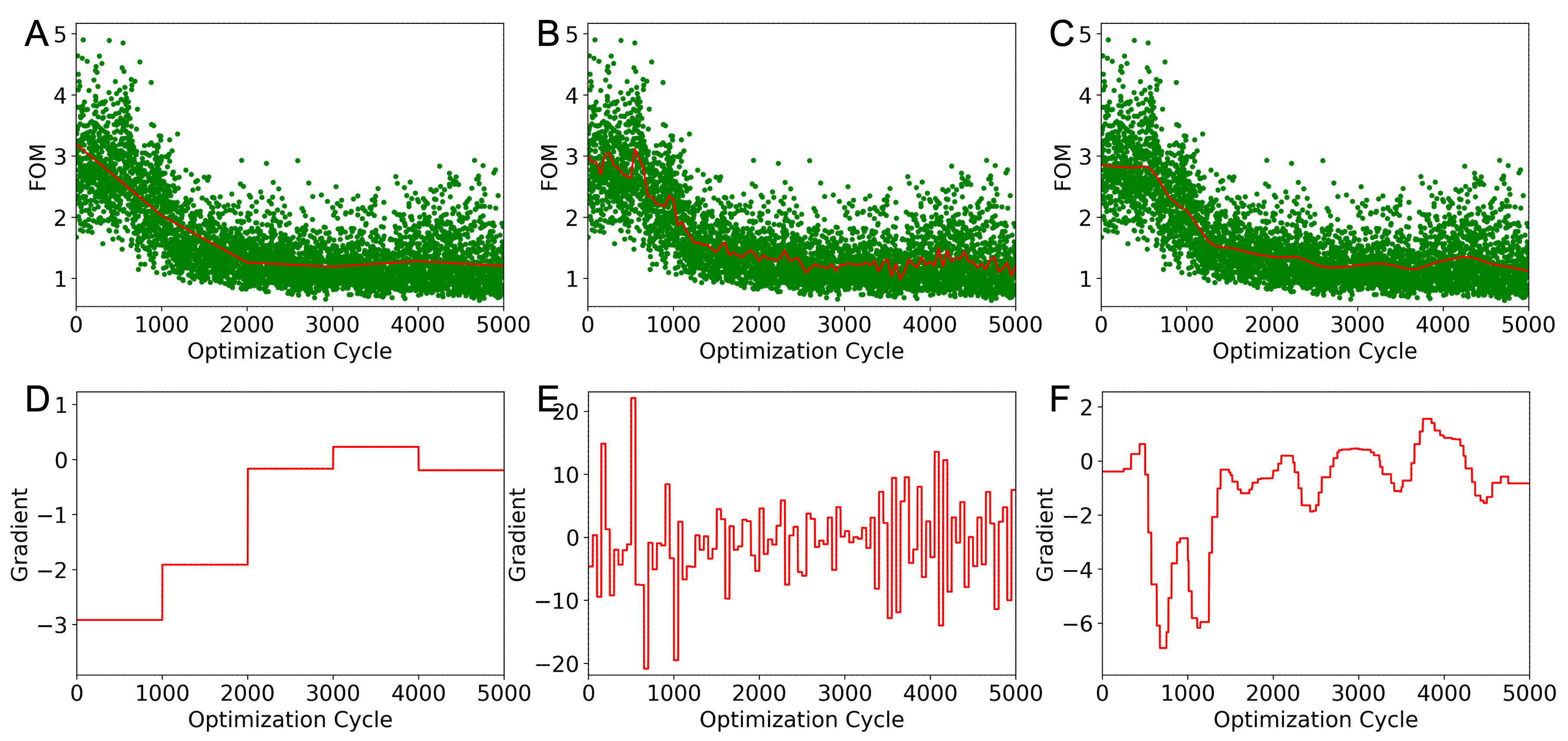}
\caption{\label{fig-S1} The analysis of FOM distributions (green dots) with piecewise linear regression (red lines). (A-C) FOM distributions and regression results after 5,000 iterations of active learning for a 120-bit system (60-layered TRC) starting optimization with 500 initial data. Piecewise linear regression is used where (A) 5 pieces and (B) 100 pieces are included. (C) Averaged piecewise linear regression is applied. (D-F) The gradient of piecewise regression lines for Figures (A-C).}
\end{figure}

Next, we apply a piecewise linear regression technique, where the regression is affected by the number of pieces, which determines the range of regression for each piece \cite{jekel2019pwlf}. Figures~\ref{fig-S1} shows that piecewise linear regression with a small number of pieces (5) cannot capture sudden changes in FOM adequately. In contrast, a large number of pieces (100) overestimates FOM distributions, making it challenging to determine the convergence point with a gradient plot (Figures 2E and S1E). Remarkably, averaged piecewise linear regression effectively captures such complex distributions, yielding reasonable regression and gradient plots (Figures~\ref{fig-S1}). The results clearly illustrate that regression captures complex FOM distributions, where FOM tends to remain at a high-value region until $\sim$500 optimization cycles before decreasing towards convergence. Hence, we adapt the averaged piecewise linear regression technique to analyze the convergence using the preset threshold (i.e., a gradient of a regression line: -3).

\subsection{Optimal Number of Initial Data}
\begin{figure}[!ht]
\centering
\includegraphics[width=1\linewidth]{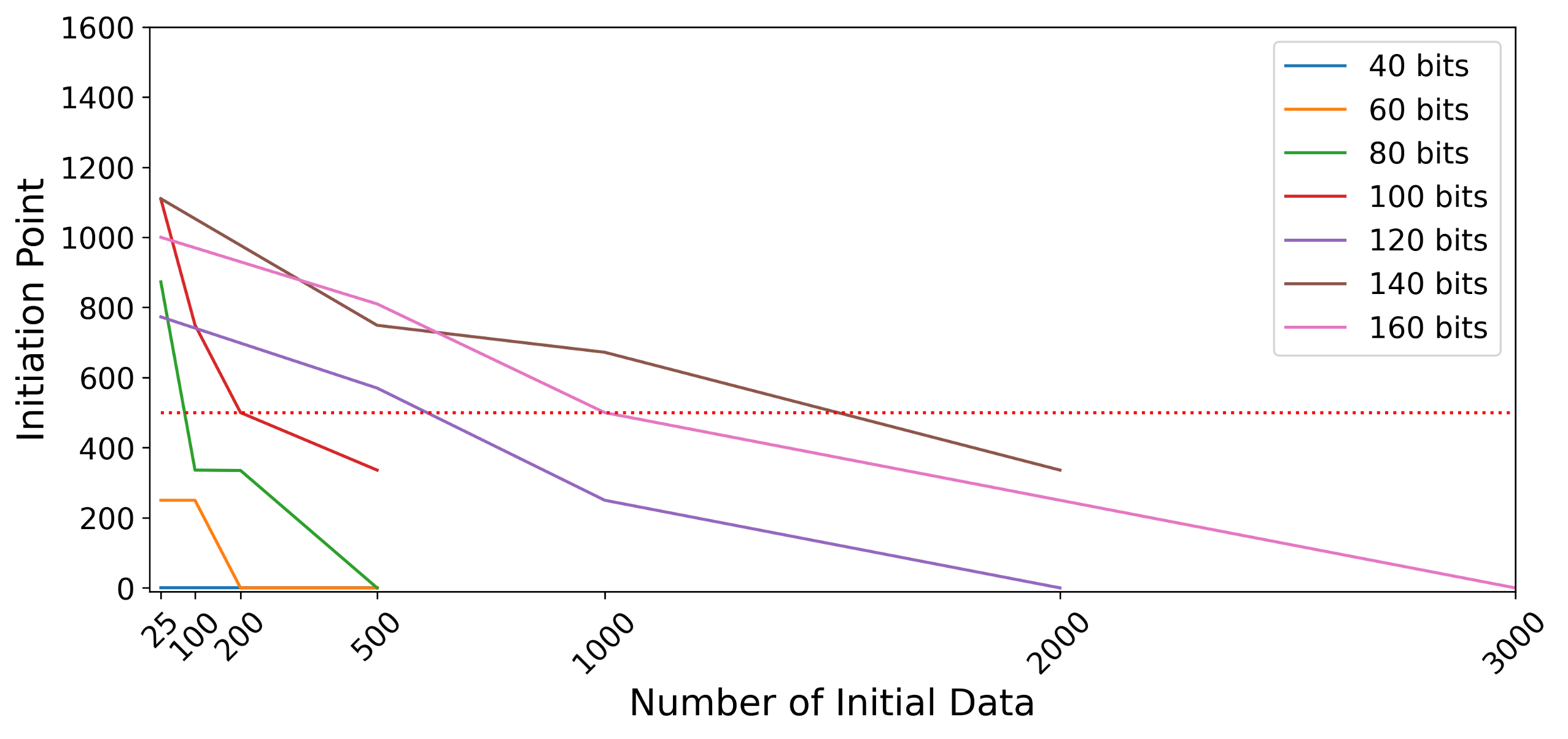}
\caption{\label{fig:fig-3} Initiation points where convergence starts as a function of the number of initial data for different design space sizes. The initiation points are determined by the predefined threshold ($-3$) to the gradients of regression plots (Figure S1), which clearly verifies faster convergence achieved by more initial data for larger systems. Note that the threshold to analyze the 160-bit system is $-2$.}
\end{figure}

Figure~\ref{fig:fig-3} shows the initiation point for convergence across different design space sizes when starting optimization with different numbers of initial data. The results demonstrate that small systems do not require a large number of initial data; for example, 40 and 60-bit systems can achieve convergence within 500 optimization cycles even with 25 initial data. On the other hand, greater numbers of initial data are required to achieve convergence for larger systems. For example, 80, 100, 120, and 140-bit systems respectively need 100, 200, 1,000, and 2,000 initial data to ensure satisfactory convergence within 500 iterations. Otherwise, active learning may require more iterations, thereby resulting in a prolonged optimization process or failure to identify optimal states, which increases computational costs for overall optimization. 

\begin{figure}[!ht]
\centering
\includegraphics[width=1.0\linewidth]{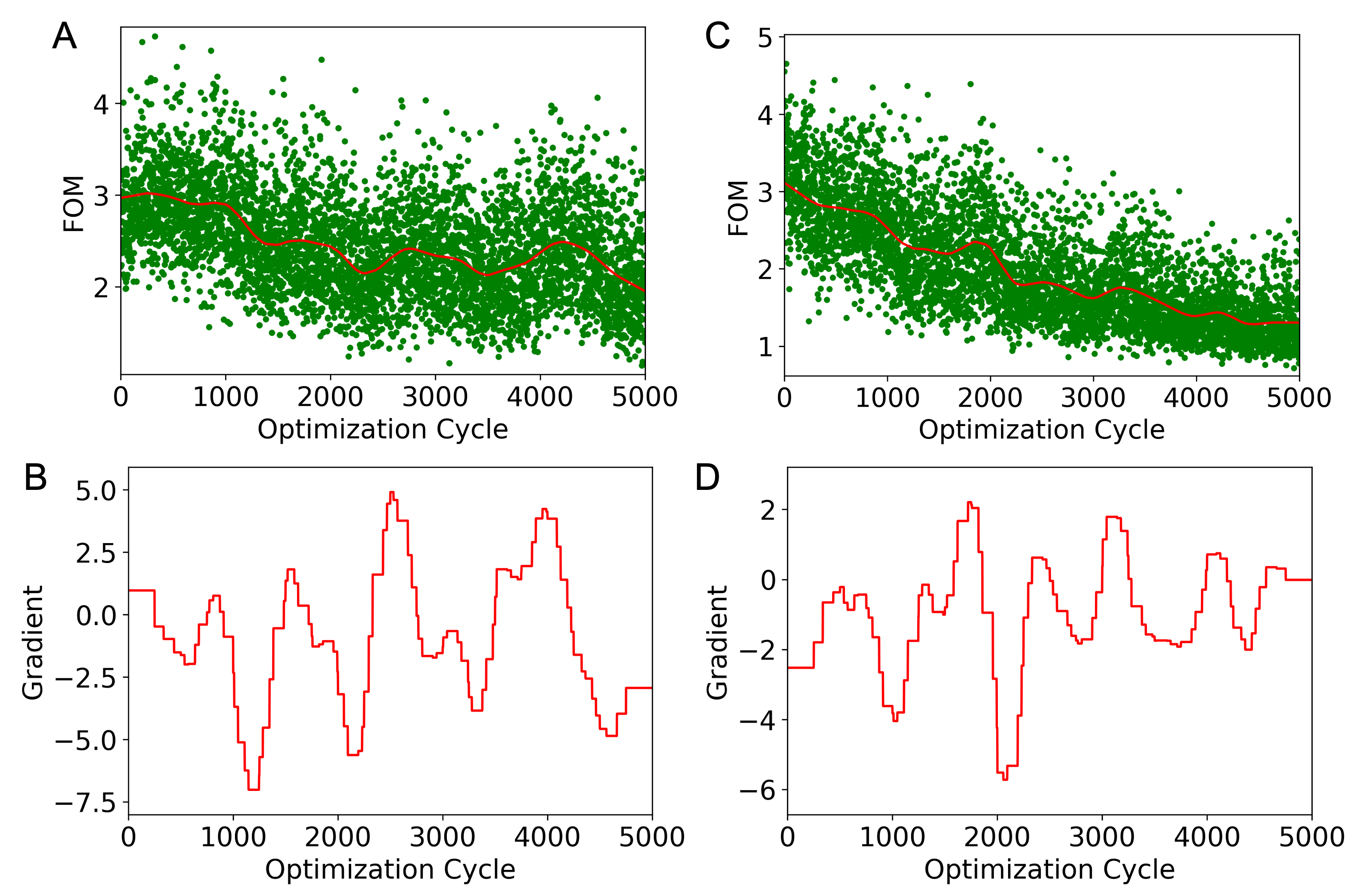}
\caption{\label{fig-S2} Optimization results after 5,000 iterations of active learning for a 160-bit system (80-layered TRC). Optimization starts with (A,B) 25 and (C,D) 3,000 initial data. (A,C) FOM distributions (green dots) and regression lines from averaged piecewise linear regression (red lines), and (B,D) corresponding gradient of the regression line.}
\end{figure}

It is worth noting that the design space of the 160-bit system is significantly large, thus it is hard to see a clear convergence pattern of FOM when starting optimization with 25 initial data (Figures~\ref{fig-S2}A,B). Conversely, FOM converges well when employing substantially larger initial data (3,000), as depicted in Figure~\ref{fig-S2}C. This means that optimization with a small number of initial data delays FOM convergence as well as often leads to failure of the overall optimization processes. 

\begin{figure}[!ht]
\centering
\includegraphics[width=0.5\linewidth]{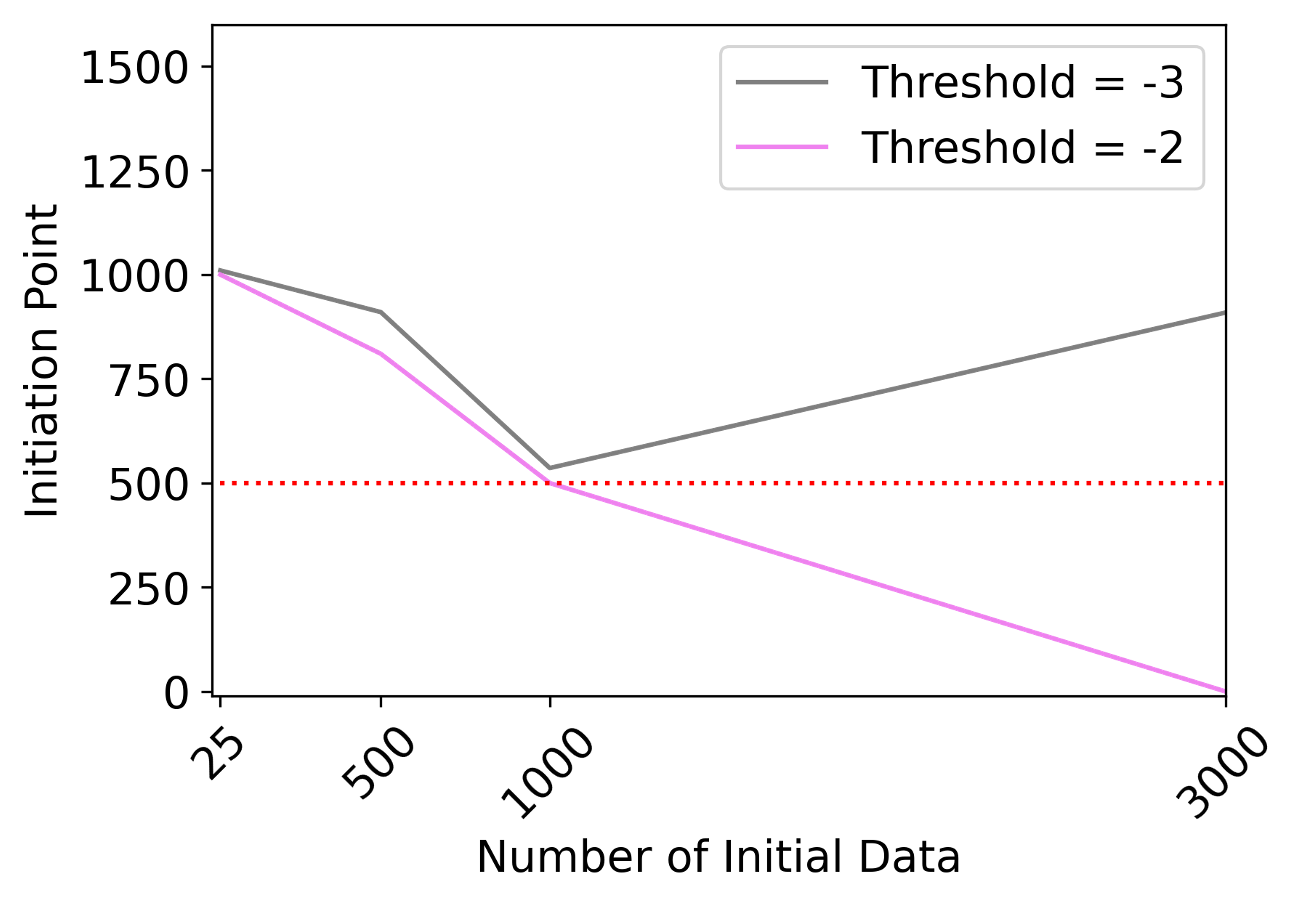}
\caption{\label{fig-S3} Initiation points where convergence starts as a function of the number of initial data for a problem size of 160 (i.e., 160-bit system / 80-layered TRC window). The initiation points are determined by the predefined threshold (-3 or -2).}
\end{figure}

Hence, optimization with an adequate number of initial data is essential for surrogate-based active learning especially when designing large systems. The absolute gradient values of the regression line are relatively small although a satisfactory convergence is observed (Figure~\ref{fig-S2}D). Hence, the initiation point for convergence is 909 if the threshold is $-3$, which is overly underestimated (Figure~\ref{fig-S2}D). Adjusting the threshold from $-3.0$ to $-2.0$ yields a more accurate determination of the initiation point, which is aligned with the observed trends in smaller systems (i.e., 40 to 140-bit systems, Figures~\ref{fig:fig-3} and \ref{fig-S3}). Therefore, it is the more proper strategy to determine the initiation point with smaller absolute threshold values for a large system. The results highlight the optimal numbers of initial data to achieve efficient and reliable convergence in optimization processes, resulting in good optimization results with reasonable computational costs.

\subsection{Optimized Functional Material}
We design TRC windows by employing this strategy, and the 60-bit system (30-layered TRC window) yields the lowest FOM (0.5027), as depicted in Figure~\ref{fig-S4}A. A binary vector representing the optimized TRC window is [11 11 11 00 01 00 11 11 10 00 00 10 11 11 01 00 01 01 11 11 11 11 01 00 00 00 10 11 11 10]. Note that this work primarily focuses on studying the convergence for different design space sizes according to different initial data. Hence, achieving a global optimal structure may require additional optimization cycles, and thus the current FOM may not represent a global minimum. Nevertheless, the presented FOM is greatly lower compared to randomly selected points. For instance, a TRC window (30-layered) composed of randomly generated structures exhibits a substantially higher FOM of 3.9389, with distinctly different optical properties to the ideal TRC window (Figures~\ref{fig-S4}). On the other hand, the optimized TRC window has a low FOM and exhibits the desired optical properties, featuring high transmission in the visible regime and low transmission in the UV and NIR regimes (Figure~\ref{fig-S4}B). Furthermore, this window has high emission in the M/LWIR regimes owing to the top polymer layer (Figure~\ref{fig-S4}C). 

\begin{figure}[!ht]
\centering
\includegraphics[width=1.0\linewidth]{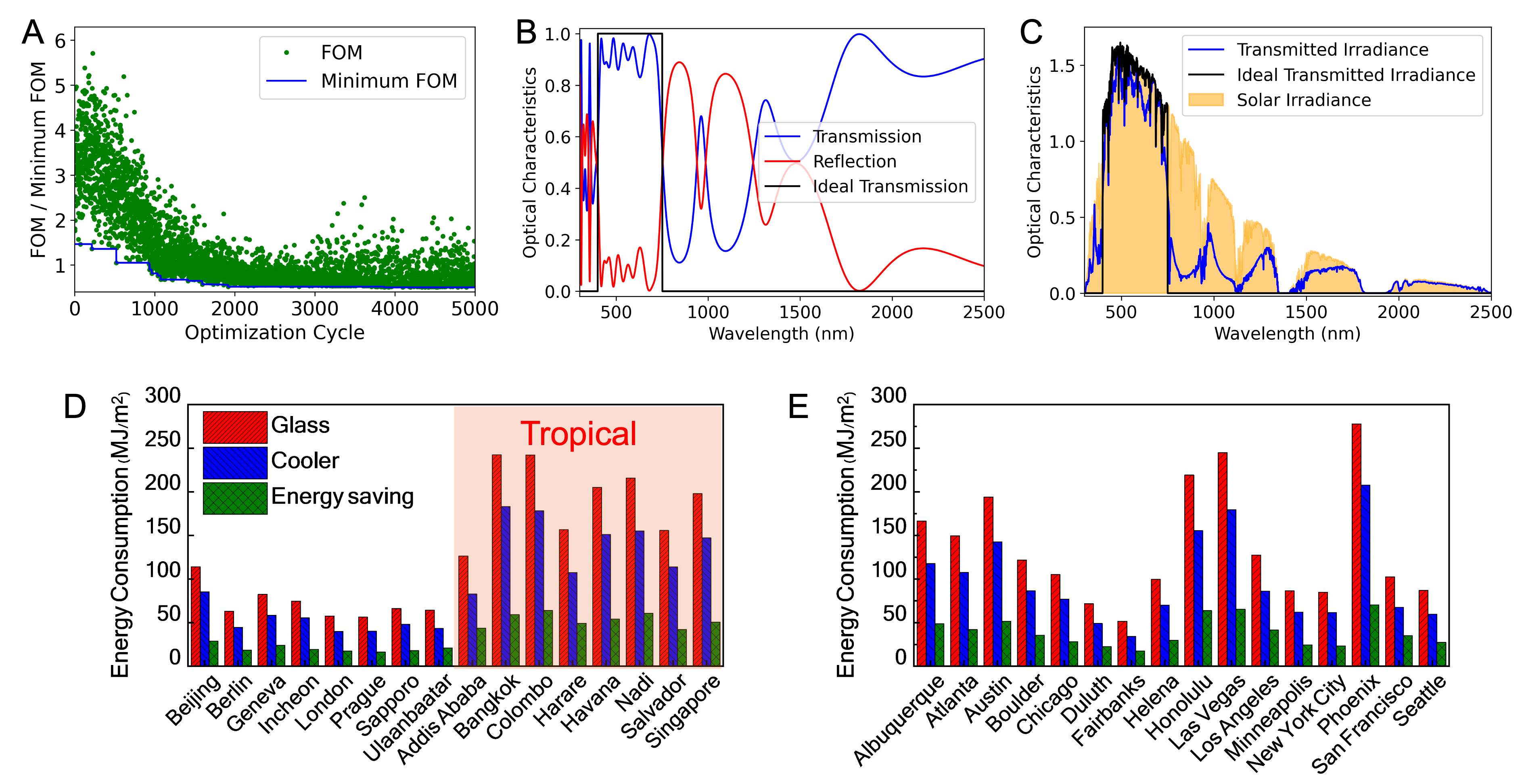}
\caption{\label{fig:fig-4} Optimization results with the surrogate-based active learning algorithm. (A) FOM distribution (green dots) and minimum FOM (blue line) after optimizing a 60-bit system (30-layered TRC) with 100 initial data. (B,C) Optical properties of the optimized TRC window. Annual energy consumption calculations for cooling in selected cities in the (D) world and (E) United States in temperate and tropical climates}
\end{figure}

\begin{figure}[!ht]
\centering
\includegraphics[width=1.0\linewidth]{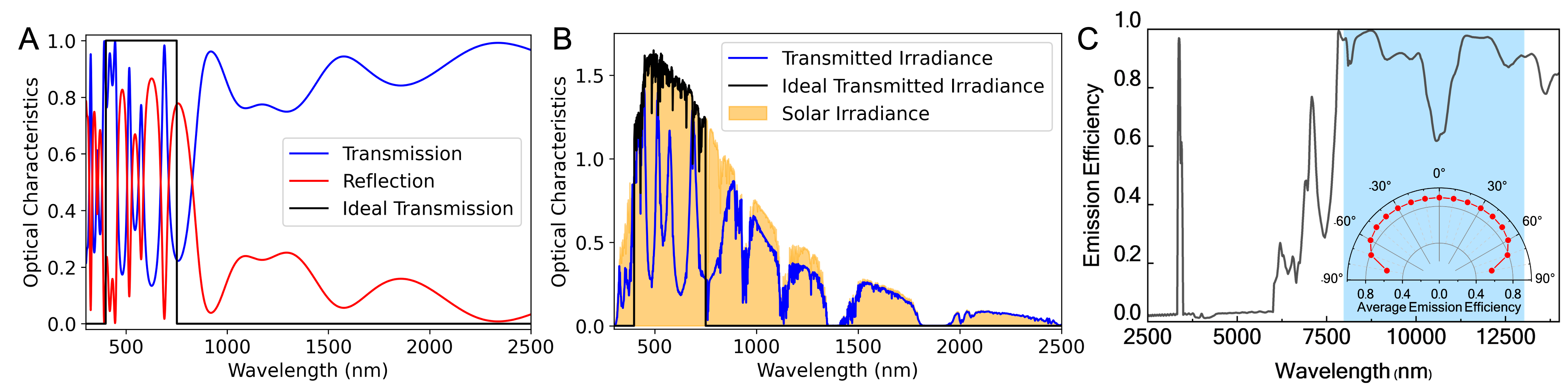}
\caption{\label{fig-S4} Randomly designed 30-layered TRC window (60-bit system). (A,B) Optical properties of the random TRC window. A binary vector representing this random TRC window is [00 00 11 01 00 11 11 00 10 11 11 01 10 00 01 01 11 11 01 01 11 10 01 10 11 11 01 00 11 10] and its FOM is 3.9389.}
\end{figure}

Consequently, the solar-weighted transmission (i.e., transmitted irradiance) of the designed TRC window closely resembles that of the ideal one, aligning with the optimization goal (Figure~\ref{fig:fig-4}). The results demonstrate that the designed TRC has a strong ability to reflect heat-generating photons while allowing visible light transmission, indicating great potential for use in building or automobile windows. To further investigate its practical applicability, we calculate energy consumption for cooling in various cities using EnergyPlus software (v9.4), by comparing scenarios with the designed TRC window or a glass window in a standard office \cite{wang2021scalable, kim2022high, kim2024wide}. 

Figures~\ref{fig:fig-4}D,E demonstrate that the TRC window requires less energy consumption for cooling compared to conventional glass windows (up to $\sim$34\% reduction), indicative of great energy-saving potential. In particular, it exhibits superior energy-saving capability in tropical climates (Figure~\ref{fig:fig-4}D). The energy calculation results indicate most cities located in temperate and tropical climates benefit from using the optimized TRC window to reduce cooling energy consumption.

\section{Conclusion}
In this work, we studied finding optimal numbers of initial data according to design space sizes to achieve reliable and efficient convergence in surrogate-based active learning. We adopted averaged piecewise linear regression to fit data by effectively modeling complex data distributions, and then we determined the initiation points where convergence starts through the predefined threshold applied to gradient plots of the regression. The results highlight the importance of leveraging more initial data to accelerate and enhance convergence for optimizing functional materials, especially for larger systems. To validate our approach, we applied it to the design of TRC windows as demonstration cases. The optimized TRC window had a low FOM, indicative of optical properties closely resembling the ideal one, which is in contrast to the randomly designed window. Consequently, the designed window showed great potential in saving cooling energy consumption by up to $\sim$34\% compared to conventional glass windows, with greater benefits observed in hot climates. Overall, this study provides insights into determining the appropriate number of initial data according to design space sizes, thereby achieving more efficient optimization results and minimizing computational costs within active learning processes.

\section{Acknowledgments}
This research used resources of the Oak Ridge Leadership Computing Facility at the Oak Ridge National Laboratory, which is supported by the Office of Science of the U.S. Department of Energy under Contract No. DE-AC05-00OR22725. This material is based upon work supported by the U.S. Department of Energy, Office of Science, National Quantum Information Science Research Centers, Quantum Science Center.
{\it Notice}: This manuscript has in part been authored by UT-Battelle, LLC under Contract No. DE-AC05-00OR22725 with the U.S. Department of Energy. The United States Government retains and the publisher, by accepting the article for publication, acknowledges that the U.S. Government retains a non-exclusive, paid up, irrevocable, world-wide license to publish or reproduce the published form of the manuscript, or allow others to do so, for U.S. Government purposes. The Department of Energy will provide public access to these results of federally sponsored research in accordance with the DOE Public Access Plan (http://energy.gov/downloads/doe-publicaccess-plan).

\printbibliography

@article{kim2022high,
  title={High-performance transparent radiative cooler designed by quantum computing},
  author={Kim, Seongmin and Shang, Wenjie and Moon, Seunghyun and Pastega, Trevor and Lee, Eungkyu and Luo, Tengfei},
  journal={ACS Energy Letters},
  volume={7},
  number={12},
  pages={4134--4141},
  year={2022},
  publisher={ACS Publications}
}

@article{kim2024quantum,
  title={Quantum annealing-aided design of an ultrathin-metamaterial optical diode},
  author={Kim, Seongmin and Park, Su-Jin and Moon, Seunghyun and Zhang, Qiushi and Hwang, Sanghyo and Kim, Sun-Kyung and Luo, Tengfei and Lee, Eungkyu},
  journal={Nano Convergence},
  volume={11},
  number={1},
  pages={1--11},
  year={2024},
  publisher={SpringerOpen}
}

@article{kim2024wide,
  title={Wide-angle spectral filter for energy-saving windows designed by quantum annealing-enhanced active learning},
  author={Kim, Seongmin and Jung, Serang and Bobbitt, Alexandria and Lee, Eungkyu and Luo, Tengfei},
  journal={Cell Reports Physical Science},
  year={2024},
  publisher={Elsevier}
}

@article{kim2023design,
  title={Design of a High-Performance Titanium Nitride Metastructure-Based Solar Absorber Using Quantum Computing-Assisted Optimization},
  author={Kim, Seongmin and Wu, Shiwen and Jian, Ruda and Xiong, Guoping and Luo, Tengfei},
  journal={ACS Applied Materials \& Interfaces},
  volume={15},
  number={34},
  pages={40606--40613},
  year={2023},
  publisher={ACS Publications}
}

@article{kim2024performance,
  title={Performance Analysis of an Optimization Algorithm for Metamaterial Design on the Integrated High-Performance Computing and Quantum Systems},
  author={Kim, Seongmin and Suh, In-Saeng},
  journal={arXiv preprint arXiv:2405.02211},
  year={2024}
}

@article{liu2020machine,
  title={Machine learning assisted materials design and discovery for rechargeable batteries},
  author={Liu, Yue and Guo, Biru and Zou, Xinxin and Li, Yajie and Shi, Siqi},
  journal={Energy Storage Materials},
  volume={31},
  pages={434--450},
  year={2020},
  publisher={Elsevier}
}

@article{himanen2019data,
  title={Data-driven materials science: status, challenges, and perspectives},
  author={Himanen, Lauri and Geurts, Amber and Foster, Adam Stuart and Rinke, Patrick},
  journal={Advanced Science},
  volume={6},
  number={21},
  pages={1900808},
  year={2019},
  publisher={Wiley Online Library}
}

@article{shang2023hybrid,
  title={Hybrid Data-Driven Discovery of High-Performance Silver Selenide-Based Thermoelectric Composites},
  author={Shang, Wenjie and Zeng, Minxiang and Tanvir, ANM and Wang, Ke and Saeidi-Javash, Mortaza and Dowling, Alexander and Luo, Tengfei and Zhang, Yanliang},
  journal={Advanced Materials},
  volume={35},
  number={47},
  pages={2212230},
  year={2023},
  publisher={Wiley Online Library}
}

@article{chen2020generative,
  title={Generative deep neural networks for inverse materials design using backpropagation and active learning},
  author={Chen, Chun-Teh and Gu, Grace X},
  journal={Advanced Science},
  volume={7},
  number={5},
  pages={1902607},
  year={2020},
  publisher={Wiley Online Library}
}

@article{wei2020genetic,
  title={Genetic algorithm-driven discovery of unexpected thermal conductivity enhancement by disorder},
  author={Wei, Han and Bao, Hua and Ruan, Xiulin},
  journal={Nano Energy},
  volume={71},
  pages={104619},
  year={2020},
  publisher={Elsevier}
}

@article{jiang2021metasurface,
  title={Metasurface based on inverse design for maximizing solar spectral absorption},
  author={Jiang, Xinpeng and Yuan, Huan and Chen, Dingbo and Zhang, Zhaojian and Du, Te and Ma, Hansi and Yang, Junbo},
  journal={Advanced Optical Materials},
  volume={9},
  number={19},
  pages={2100575},
  year={2021},
  publisher={Wiley Online Library}
}

@article{wang2020machine,
  title={Machine learning approaches for thermoelectric materials research},
  author={Wang, Tian and Zhang, Cheng and Snoussi, Hichem and Zhang, Gang},
  journal={Advanced Functional Materials},
  volume={30},
  number={5},
  pages={1906041},
  year={2020},
  publisher={Wiley Online Library}
}

@article{ha2023rapid,
  title={Rapid inverse design of metamaterials based on prescribed mechanical behavior through machine learning},
  author={Ha, Chan Soo and Yao, Desheng and Xu, Zhenpeng and Liu, Chenang and Liu, Han and Elkins, Daniel and Kile, Matthew and Deshpande, Vikram and Kong, Zhenyu and Bauchy, Mathieu and others},
  journal={Nature Communications},
  volume={14},
  number={1},
  pages={5765},
  year={2023},
  publisher={Nature Publishing Group UK London}
}

@article{ma2021deep,
  title={Deep learning for the design of photonic structures},
  author={Ma, Wei and Liu, Zhaocheng and Kudyshev, Zhaxylyk A and Boltasseva, Alexandra and Cai, Wenshan and Liu, Yongmin},
  journal={Nature Photonics},
  volume={15},
  number={2},
  pages={77--90},
  year={2021},
  publisher={Nature Publishing Group UK London}
}

@article{pestourie2023physics,
  title={Physics-enhanced deep surrogates for partial differential equations},
  author={Pestourie, Rapha{\"e}l and Mroueh, Youssef and Rackauckas, Chris and Das, Payel and Johnson, Steven G},
  journal={Nature Machine Intelligence},
  volume={5},
  number={12},
  pages={1458--1465},
  year={2023},
  publisher={Nature Publishing Group UK London}
}

@article{wilson2021machine,
  title={Machine learning framework for quantum sampling of highly constrained, continuous optimization problems},
  author={Wilson, Blake A and Kudyshev, Zhaxylyk A and Kildishev, Alexander V and Kais, Sabre and Shalaev, Vladimir M and Boltasseva, Alexandra},
  journal={Applied Physics Reviews},
  volume={8},
  number={4},
  year={2021},
  publisher={AIP Publishing}
}

@article{kitai2020designing,
  title={Designing metamaterials with quantum annealing and factorization machines},
  author={Kitai, Koki and Guo, Jiang and Ju, Shenghong and Tanaka, Shu and Tsuda, Koji and Shiomi, Junichiro and Tamura, Ryo},
  journal={Physical Review Research},
  volume={2},
  number={1},
  pages={013319},
  year={2020},
  publisher={APS}
}

@article{pastorello2019quantum,
  title={Quantum annealing learning search for solving QUBO problems},
  author={Pastorello, Davide and Blanzieri, Enrico},
  journal={Quantum Information Processing},
  volume={18},
  number={10},
  pages={303},
  year={2019},
  publisher={Springer}
}

@article{lye2021iterative,
  title={Iterative surrogate model optimization (ISMO): An active learning algorithm for PDE constrained optimization with deep neural networks},
  author={Lye, Kjetil O and Mishra, Siddhartha and Ray, Deep and Chandrashekar, Praveen},
  journal={Computer Methods in Applied Mechanics and Engineering},
  volume={374},
  pages={113575},
  year={2021},
  publisher={Elsevier}
}

@article{pestourie2020active,
  title={Active learning of deep surrogates for PDEs: application to metasurface design},
  author={Pestourie, Rapha{\"e}l and Mroueh, Youssef and Nguyen, Thanh V and Das, Payel and Johnson, Steven G},
  journal={npj Computational Materials},
  volume={6},
  number={1},
  pages={164},
  year={2020},
  publisher={Nature Publishing Group UK London}
}

@article{kapadia2024active,
  title={Active-learning-driven surrogate modeling for efficient simulation of parametric nonlinear systems},
  author={Kapadia, Harshit and Feng, Lihong and Benner, Peter},
  journal={Computer Methods in Applied Mechanics and Engineering},
  volume={419},
  pages={116657},
  year={2024},
  publisher={Elsevier}
}

@article{ren2021survey,
  title={A survey of deep active learning},
  author={Ren, Pengzhen and Xiao, Yun and Chang, Xiaojun and Huang, Po-Yao and Li, Zhihui and Gupta, Brij B and Chen, Xiaojiang and Wang, Xin},
  journal={ACM computing surveys (CSUR)},
  volume={54},
  number={9},
  pages={1--40},
  year={2021},
  publisher={ACM New York, NY}
}

@article{molesky2018inverse,
  title={Inverse design in nanophotonics},
  author={Molesky, Sean and Lin, Zin and Piggott, Alexander Y and Jin, Weiliang and Vuckovi{\'c}, Jelena and Rodriguez, Alejandro W},
  journal={Nature Photonics},
  volume={12},
  number={11},
  pages={659--670},
  year={2018},
  publisher={Nature Publishing Group UK London}
}

@article{zunger2018inverse,
  title={Inverse design in search of materials with target functionalities},
  author={Zunger, Alex},
  journal={Nature Reviews Chemistry},
  volume={2},
  number={4},
  pages={0121},
  year={2018},
  publisher={Nature Publishing Group UK London}
}

@article{kusne2020fly,
  title={On-the-fly closed-loop materials discovery via Bayesian active learning},
  author={Kusne, A Gilad and Yu, Heshan and Wu, Changming and Zhang, Huairuo and Hattrick-Simpers, Jason and DeCost, Brian and Sarker, Suchismita and Oses, Corey and Toher, Cormac and Curtarolo, Stefano and others},
  journal={Nature communications},
  volume={11},
  number={1},
  pages={5966},
  year={2020},
  publisher={Nature Publishing Group UK London}
}

@article{li2019radiative,
  title={A radiative cooling structural material},
  author={Li, Tian and Zhai, Yao and He, Shuaiming and Gan, Wentao and Wei, Zhiyuan and Heidarinejad, Mohammad and Dalgo, Daniel and Mi, Ruiyu and Zhao, Xinpeng and Song, Jianwei and others},
  journal={Science},
  volume={364},
  number={6442},
  pages={760--763},
  year={2019},
  publisher={American Association for the Advancement of Science}
}

@article{wang2021scalable,
  title={Scalable thermochromic smart windows with passive radiative cooling regulation},
  author={Wang, Shancheng and Jiang, Tengyao and Meng, Yun and Yang, Ronggui and Tan, Gang and Long, Yi},
  journal={Science},
  volume={374},
  number={6574},
  pages={1501--1504},
  year={2021},
  publisher={American Association for the Advancement of Science}
}

@article{zhu2021subambient,
  title={Subambient daytime radiative cooling textile based on nanoprocessed silk},
  author={Zhu, Bin and Li, Wei and Zhang, Qian and Li, Duo and Liu, Xin and Wang, Yuxi and Xu, Ning and Wu, Zhen and Li, Jinlei and Li, Xiuqiang and others},
  journal={Nature nanotechnology},
  volume={16},
  number={12},
  pages={1342--1348},
  year={2021},
  publisher={Nature Publishing Group UK London}
}

@article{dang2022ultrathin,
  title={An ultrathin transparent radiative cooling photonic structure with a high NIR reflection},
  author={Dang, Saichao and Wang, Xiaojia and Ye, Hong},
  journal={Advanced Materials Interfaces},
  volume={9},
  number={30},
  pages={2201050},
  year={2022},
  publisher={Wiley Online Library}
}

@article{kim2021visibly,
  title={Visibly transparent radiative cooler under direct sunlight},
  author={Kim, Minkyung and Lee, Dasol and Son, Soomin and Yang, Younghwan and Lee, Heon and Rho, Junsuk},
  journal={Advanced Optical Materials},
  volume={9},
  number={13},
  pages={2002226},
  year={2021},
  publisher={Wiley Online Library}
}

@article{volpe2023integration,
  title={Integration of simulated quantum annealing in parallel tempering and population annealing for heterogeneous-profile QUBO exploration},
  author={Volpe, Deborah and Cirillo, Giovanni Amedeo and Zamboni, Maurizio and Turvani, Giovanna},
  journal={IEEE Access},
  volume={11},
  pages={30390--30441},
  year={2023},
  publisher={IEEE}
}

@article{kim2021comparison,
  title={Comparison between multiple regression analysis, polynomial regression analysis, and an artificial neural network for tensile strength prediction of BFRP and GFRP},
  author={Kim, Younghwan and Oh, Hongseob},
  journal={Materials},
  volume={14},
  number={17},
  pages={4861},
  year={2021},
  publisher={MDPI}
}

@article{yang2019piecewise,
  title={Piecewise linear regression based on plane clustering},
  author={Yang, Xubing and Yang, Hongxin and Zhang, Fuquan and Zhang, Li and Fan, Xijian and Ye, Qiaolin and Fu, Liyong},
  journal={IEEE Access},
  volume={7},
  pages={29845--29855},
  year={2019},
  publisher={IEEE}
}

@article{gelman2019high,
  title={Why high-order polynomials should not be used in regression discontinuity designs},
  author={Gelman, Andrew and Imbens, Guido},
  journal={Journal of Business \& Economic Statistics},
  volume={37},
  number={3},
  pages={447--456},
  year={2019},
  publisher={Taylor \& Francis}
}

@article{jekel2019pwlf,
  title={pwlf: a python library for fitting 1D continuous piecewise linear functions},
  author={Jekel, Charles F and Venter, Gerhard},
  journal={URL: https://github. com/cjekel/piecewise\_linear\_fit\_py},
  year={2019}
}

@article{hen2016quantum,
  title={Quantum annealing for constrained optimization},
  author={Hen, Itay and Spedalieri, Federico M},
  journal={Physical Review Applied},
  volume={5},
  number={3},
  pages={034007},
  year={2016},
  publisher={APS}
}

@article{kim2025distributed,
  title={Distributed Variational Quantum Algorithm with Many-qubit for Optimization Challenges},
  author={Kim, Seongmin and Suh, In-Saeng},
  journal={arXiv preprint arXiv:2503.00221},
  year={2025}
}

@article{kim2024distributed,
  title={Distributed quantum approximate optimization algorithm on integrated high-performance computing and quantum computing systems for large-scale optimization},
  author={Kim, Seongmin and Luo, Tengfei and Lee, Eungkyu and Suh, In-Saeng},
  journal={arXiv preprint arXiv:2407.20212},
  year={2024}
}

@String{Computing = "Computing" }

@String{Computer = "{IEEE} Computer" }

@String{Springer = "Springer-Verlag" }

\end{document}